\documentclass[aps,prd,floatfix,superscriptaddress,preprintnumbers]{revtex4}
\usepackage{amssymb}
\usepackage{amsmath}
\usepackage{amsfonts}
\usepackage{epsfig}             
\usepackage{graphicx}
\usepackage{tabularx}
\usepackage{color}
\newcommand{\eq}{\begin{eqnarray}}
\newcommand{\en}{\end{eqnarray}}
\newcommand{\la}{\langle}
\newcommand{\ra}{\rangle}
\newcommand{\bfq}{{\bf q}_{\perp}}
\newcommand{\bfk}{{\bf k}_{\perp}}

\begin{document}

\preprint{SLAC-PUB-17698}

\title{Heavy quark contribution to the 
  electromagnetic properties of the nucleon} 

\author{Stanley J. Brodsky} 
\affiliation{SLAC National Accelerator Laboratory, 
Stanford University, Stanford, CA 94309, USA}
\author{Valery E. Lyubovitskij}
\affiliation{Institut f\"ur Theoretische Physik,
Universit\"at T\"ubingen,
Kepler Center for Astro and Particle Physics,
Auf der Morgenstelle 14, D-72076 T\"ubingen, Germany}
\affiliation{Departamento de F\'\i sica y Centro Cient\'\i fico
Tecnol\'ogico de Valpara\'\i so-CCTVal, Universidad T\'ecnica
Federico Santa Mar\'\i a, Casilla 110-V, Valpara\'\i so, Chile}
\affiliation{Millennium Institute for Subatomic Physics at
the High-Energy Frontier (SAPHIR) of ANID, \\
Fern\'andez Concha 700, Santiago, Chile}
\author{Ivan Schmidt}
\affiliation{Departamento de F\'\i sica y Centro Cient\'\i fico
Tecnol\'ogico de Valpara\'\i so-CCTVal, Universidad T\'ecnica
Federico Santa Mar\'\i a, Casilla 110-V, Valpara\'\i so, Chile}

\begin{abstract}

Quantum chromodynamics (QCD) predicts the existence of both nonperturbative 
intrinsic and perturbative extrinsic heavy quark contributions to  
the fundamental structure of hadrons.  The existence of intrinsic charm at the
3-standard-deviation level in the proton has recently been established
from structure function measurements by the NNPDF Collaboration.
Here we revisit the physics of intrinsic heavy quarks using 
{\it light-front holographic QCD (LFHQCD)} -- a novel comprehensive approach
to hadron structure which provides detailed predictions for dynamical
properties of the hadrons, such as form factors, distribution
amplitudes, structure functions, etc. 
We will extend this nonperturbative light-front QCD approach to study
the heavy quark-antiquark contribution to the electromagnetic properties
of nucleon. Our framework is based on a study of the eigenfunctions of
the QCD light-front Hamiltonian, the frame-independent light-front wave
functions (LFWFs) underlying hadron dynamics.
We analyze the heavy quark content in the proton,
induced either directly by the nonperturbative $|uud+Q\bar Q\ra$
Fock state or by the $|uud+g\ra$ Fock state, where the gluon splits 
into a heavy quark-antiquark pair.
The specific form of these LFWFs are derived from LFHQCD.
Using these LFWFs, we construct light-front representations for the
heavy quark-antiquark asymmetry, the
electromagnetic form factors of nucleons induced by heavy quarks, 
including their magnetic moments and radii. 

\end{abstract}

\maketitle 

One of the rigorous predictions of quantum chromodynamics (QCD) is 
the existence of both nonperturbative intrinsic and perturbative 
extrinsic heavy quark contents of the
nucleon~\cite{Brodsky:1981se,Brodsky:1980pb}.
The existence of intrinsic charm at the 3-standard-deviation level
has recently been established from a comprehensive analysis of
structure function measurements
by the NNPDF Collaboration~\cite{Ball:2022qks}.

The full heavy quark parton distribution (PDF) 
is given by the sum of intrinsic (in) and extrinsic (ex) components: 
$Q(x) = Q_{\rm in}(x) + Q_{\rm ex}(x)$, 
where $Q=c,b$ and $x=\frac{k^+}{p^+}$ is the light-front variable. 
It was assumed in Refs.~\cite{Brodsky:1981se,Brodsky:1980pb} 
that the intrinsic contribution is induced by the twist-5 Fock
state $|uud+Q\bar Q\ra$. For a review see, e.g. Ref.~\cite{Brodsky:2015fna}. 
The light-front wave functions (LFWFs) are the Fock state projections of
the QCD light-front Hamiltonian. 

\begin{figure}[hb]
\begin{center}
\epsfig{figure=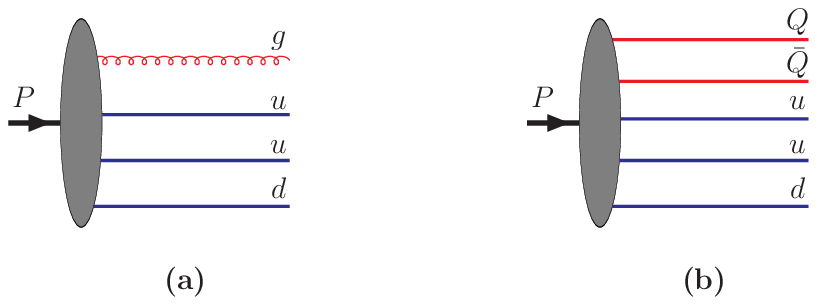,scale=.6}
\caption{Proton Fock states producing heavy quark-antiquark 
  contributions: (a) $|uud+g\ra$ state and (b) $|uud+Q\bar Q\ra$ state.
\label{fig1}}
\end{center}
\end{figure}

We will use the light-front QCD approach for the study of a heavy
quark-antiquark contribution to the electromagnetic properties
of nucleon. Our framework is based on the boost-invariant LFWFs,
effectively describing the heavy quark-antiquark
content in the nucleon. We will focus on two nonperturbative Fock 
states $|uud+Q\bar Q\ra$ and $|uud+g\ra$ (see Fig.~\ref{fig1}). 
The latter gives rise to the heavy quark-antiquark contribution via gluon
splitting into a $Q\bar Q$ pair. The analytic form of these LFWFs is predicted by 
light-front holographic QCD (LFHQCD)~\cite{Brodsky:2014yha}-\cite{Lyubovitskij:2020otz} 
which we will apply to the heavy quark contribution to the 
heavy quark-antiquark asymmetry and the electromagnetic properties
of nucleons: form factors, magnetic moment, and radii. 

We will derive the LFWFs $\psi^{\lambda_N}_{Q; \lambda_Q}(x,\bfk)$ 
and $\psi^{\lambda_N}_{\bar Q; \lambda_{\bar Q}}(x,\bfk)$  
describing heavy quark-antiquark contribution in the proton
with specific helicities
for the proton $\lambda_N = \uparrow$ and $\downarrow$ 
and for the struck heavy quark $\lambda_Q = \pm\frac{1}{2}$
or antiquark $\lambda_{\bar Q} = \pm\frac{1}{2}$. 

The LFWFs, which determine the heavy struck quark distribution 
$\psi^{\lambda_N}_{Q; \lambda_Q}(x,\bfk)$,  
are listed as: 
\eq
\psi^{\uparrow}_{Q; + \frac{1}{2}}(x,\bfk) &=& \varphi(x,\bfk) 
\nonumber\\
\psi^{\uparrow}_{Q; - \frac{1}{2}}(x,\bfk) &=&
- \dfrac{k^1 + i k^2}{\kappa} \, \varphi(x,\bfk) 
\,, \nonumber\\
\psi^{\downarrow}_{Q; + \frac{1}{2}}(x,\bfk) &=&
- \Big[\psi^{\uparrow}_{Q; - \frac{1}{2}}(x,\bfk)\Big]^\dagger =
\dfrac{k^1 - i k^2}{\kappa} \, \varphi(x,\bfk) 
\,, \nonumber\\
\psi^{\downarrow}_{Q; - \frac{1}{2}}(x,\bfk) &=&
\Big[\psi^{\uparrow}_{Q; + \frac{1}{2}}(x,\bfk)\Big]^\dagger =
\varphi(x,\bfk) \,,
\en
where
\eq
\varphi(x,\bfk) = \dfrac{2 \pi \sqrt{2}}{\kappa} \, 
\sqrt{Q_{\rm in}(x)} \, \exp\Big[-\dfrac{\bfk^2}{2\kappa^2}\Big] \,,
\en
$\kappa = 500$ MeV is the scale dilaton parameter,  
and 
$Q_{\rm in}(x)$ is the intrinsic heavy quark PDF,  
which is expressed in terms of the derived LFWFs as 
\eq 
Q_{\rm in}(x) &=&
\int \frac{d^2\bfk}{16\pi^3} \,
\biggl[ |\psi^{\uparrow}_{Q; + \frac{1}{2}}(x,\bfk)|^2
      + |\psi^{\uparrow}_{Q;- \frac{1}{2}}(x,\bfk)|^2
\biggr] \nonumber\\
&=&
\int \frac{d^2\bfk}{16\pi^3} \,
\biggl[ |\psi^{\downarrow}_{Q; + \frac{1}{2}}(x,\bfk)|^2
      + |\psi^{\downarrow}_{Q; - \frac{1}{2}}(x,\bfk)|^2
\biggr] \,.
\en
We can assume that the probability of intrinsic heavy quarks
in the proton scales as $\frac{1}{m^2_Q}$~\cite{Brodsky:2015fna},
a result that comes from
a two-gluon intermediate state perturbative contribution,
in contrast to the logarithmic dependence of the extrinsic
contribution generated by pQCD evolution. 

For the $Q_{\rm in}(x)$ we can use
the result obtained in LF QCD~\cite{Brodsky:1981se}: 
\eq\label{qin}
Q_{\rm in}(x) = N_{Q_{\rm in}} \, x^2 \, \biggl[ (1-x) (1 + 10 x + x^2)
+ 6 x (1+x) \log(x) \biggr] \,,
\en
where $N_{Q_{\rm in}}$ is the normalization constant, fixed from data. 
We have assumed that the normalization constant of the intrinsic
heavy quark distribution, $N_{c_{\rm in}}$ in Eq.~(\ref{qin}), is equal to 6,
corresponding to a 1\% intrinsic charm contribution to the proton PDF.
This choice was motivated by an estimate of the magnitude
of the diffractive production of the $\Lambda_c$ baryon in the
$pp \to p \Lambda_c X$ reaction~\cite{Brodsky:1981se}, which is consistent
with the MIT bag-model estimate~\cite{Donoghue:1977qp} of the probability
of finding a five-quark $|uud+Q\bar Q\ra$ configuration the nucleon
at the order of 1\%$-$2\%. In the case of the bottom distribution, 
$N_{b_{\rm in}} = 6 \, (m_c/m_b)^2$~\cite{Brodsky:2015fna}.

The LFWFs with a heavy struck antiquark 
$\psi^{\lambda_N}_{\lambda_{\bar Q}}(x,\bfk)$ 
are obtained from the LFWFs for a heavy struck quark,
upon replacement of the heavy quark PDF by
the heavy antiquark PDF, as 
\eq 
Q_{\rm in}(x) \to \bar Q_{\rm in}(x) = 
N \, (1-x) \, Q_{\rm in}(x)  \,. 
\en
We have assumed that the $\bar Q_{\rm in}(x)$ has an 
at least $(1-x)$ falloff at large $x$ in comparison
with $Q_{\rm in}(x)$;  i.e., $\bar Q_{\rm in}(x) \sim (1-x)^6$ 
at $x \to 1$. The normalization constant $N$ will be fixed using 
an explicit form of the $Q_{\rm in}(x)$.

Following Ref.~\cite{Sufian:2020coz}, we  introduce the asymmetric 
heavy-antiheavy quark distribution function
$Q_{\rm asym}(x) = Q_{\rm in}(x) - \bar Q_{\rm in}(x)$.   
Using the explicit form of $Q_{\rm in}(x)$~(\ref{qin}), 
the condition that the zero moments of the 
$Q_{\rm in}(x)$ and $\bar Q_{\rm in}(x)$ PDFs should be equal,
or that the zero moment of the asymmetry PDF $Q_{\rm asym}(x)$
should vanish, gives us that $N=7/5$ and
\eq 
\bar Q_{\rm in}(x) = \dfrac{7}{5} (1-x) \, Q_{\rm in}(x) \,. 
\en 
Finally, the LFWFs with a heavy struck antiquark 
$\psi^{\lambda_N}_{\bar Q; \lambda_{\bar Q}}(x,\bfk)$ are related
to the corresponding LFWFs with heavy struck quark as: 
\eq
\psi^{\lambda_N}_{\bar Q; \lambda_{\bar Q}}(x,\bfk) = 
\sqrt{\dfrac{7}{5} (1-x)} \, 
\psi^{\lambda_N}_{Q; \lambda_Q}(x,-\bfk)  \,. 
\en 

We can also consider
the sum of the heavy quark and antiquark PDFs  
\eq 
Q^+(x) = Q_{\rm in}(x) + \bar Q_{\rm in}(x)  \,, 
\en
and the fraction of the proton momentum $[Q]$ carried by heavy quark and antiquark
\eq
[Q] =  \int\limits_0^1 dx \, x \, Q^+ (x)
\en
which were recently extracted by
the NNPDF Collaboration~\cite{Ball:2022qks}.

\begin{figure}
\begin{center}
\epsfig{figure=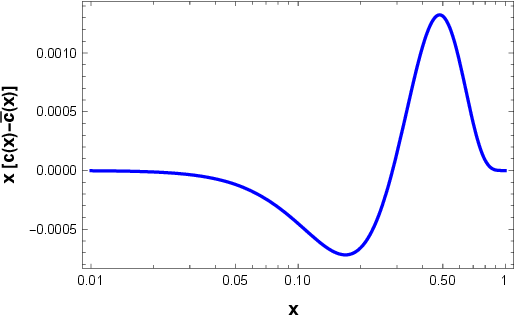,scale=.75}
\hspace*{1cm}
\epsfig{figure=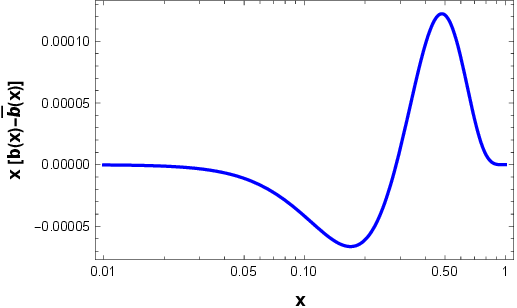,scale=.75}
\end{center}
\vspace*{-.48cm}
\noindent
\caption{Plots of the 
$x c_{\rm asym}(x)$ (left panel) and 
$x b_{\rm asym}(x)$ (right panel) asymmetric distributions. 
\label{fig2}} 
\end{figure}

In Fig.~\ref{fig2} we present numerical results for 
the asymmetric heavy-antiheavy quark distribution functions 
$Q_{\rm asym}(x)$ for charm and bottom quarks. 
We note that our results for the intrinsic charm quark 
asymmetry $c_{\rm asym}(x)$ are in very good agreement with the predictions 
in Ref.~\cite{Sufian:2020coz} based on lattice gauge theory.
It is also interesting to compare the
LFHQCD prediction for the first moment of the asymmetry: 
\eq 
\la x \ra_{Q-\bar Q} = \int\limits_0^1 dx \, x \, Q_{\rm asym}(x)
= \dfrac{1}{3500} 
\en 
with the constraint on the  charm-anticharm asymmetry
in the nucleon obtained from lattice QCD. 
Our prediction for the $\la x \ra_{c-\bar c} = 0.00029$ 
is in order of magnitude  agreement with the prediction 
of Ref.~\cite{Sufian:2020coz} $\la x \ra_{c-\bar c} = 0.00047(15)$. 
Our result for the first moment of the bottom quark asymmetry is suppressed
by a factor $(m_c/m_b)^2$ due to the corresponding ratio of the normalization
constants $N_{b_{\rm in}}/N_{c_{\rm in}} = (m_c/m_b)^2$. This dependence on
the mass of the heavy quark reflects the twist of the operators
controlling the probability for intrinsic heavy quarks in hadrons.
For the heavy quark masses, we have used the central values from the 
Particle Data Group~\cite{PDG22}: $m_c = 1.27$ GeV and
$m_b = 4.18$ GeV. This gives $\la x \ra_{b-\bar b} = 0.000026$. 

\begin{figure}[hb]
\begin{center}
\epsfig{figure=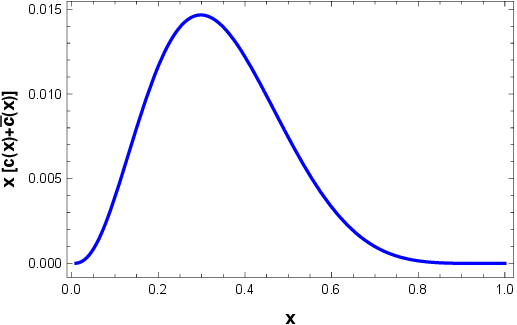,scale=.75}
\hspace*{1cm}
\epsfig{figure=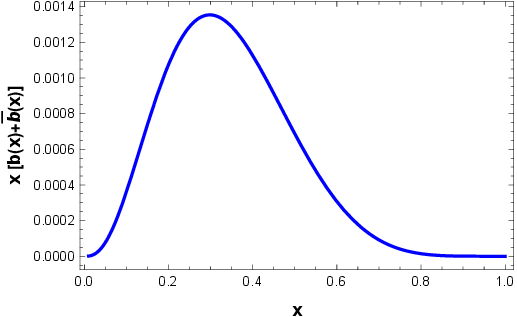,scale=.75}
\end{center}
\vspace*{-.48cm}
\noindent
\caption{Plots of the 
$x c^+(x)$ (left panel) and 
  $x b^+(x)$ (right panel) symmetric distributions. In the case of charm quarks, we 
  present a comparison with results of the NNPDF Collaboration~\cite{Ball:2022qks}
  and vary our normalization constant in the region $N_{c_{\rm in}} = 6 \pm 3$. 
\label{fig3}} 
\end{figure}

In Fig.~\ref{fig3}, we present numerical results for 
the symmetric heavy-antiheavy quark distribution functions
$x Q^+(x)$, for both charm and bottom quarks. 
One should stress that our results for the charm quark 
are in very good agreement with the empirical results obtained by
the NNPDF Collaboration~\cite{Ball:2022qks}.

In the case of the charm quark, we assume the normalization 
constant $N_{c_{\rm in}} = 6 \pm 3$, which corresponds to a variation
of the intrinsic charm  probability
in the interval $(1 \pm 0.5)\%$ to the proton PDF. Our central curve
corresponds to the central value of $N_{c_{\rm in}} = 6$; 
the shaded band corresponds to variation of $N_{c_{\rm in}}$ from 3 to 9.
Increasing $N_{c_{\rm in}}$ leads to increasing the normalization  for $x Q^+(x)$. 
The corresponding results of Ref.~\cite{Ball:2022qks}, shown as a 
central value curve and shaded band, take into account
different flavor number schemes, different orders in the $\alpha_s$,
and uncertainties of the empirical extraction.   
The main difference with our analysis is
that at $x=0$ and small $x$ our results are consistent with the physical meaning
of the PDFs: They are distributions, and must be positively defined quantities
(both for heavy quark and antiquark), whereas the NNPDF results allow for
negative distributions, due to the uncertainties mentioned above. 
  
It is interesting
to compare the predictions of our approach and NNPDF Collaboration for the fraction
of the proton momentum $[Q]$. Our prediction for the central value $N_{c_{\rm in}} = 6$
is $[Q]=0.54\%$ [with the variation of $N_{c_{\rm in}}$ we obtain $[Q]=(0.54 \pm 0.27)\%$].  
This is in good agreement with the NNPDF result: $(0.62 \pm  0.28)\%$ in the case when
only PDF uncertainties have been included and $(0.62 \pm 0.61)\%$ in the case when
the effect of missing higher-order uncertainties have also been included.

As the next step, we calculate the intrinsic contribution of heavy quarks to the nucleon
electromagnetic form factors. In LF QCD, the Dirac and Pauli  electromagnetic form
factors of the nucleon are identified with the spin-conserving and spin-flip current
matrix elements and are calculated using the Drell-Yan-West (DYW)~\cite{Drell:1969km}  
and Brodsky-Drell (BD)~\cite{Brodsky:1980zm} formulas:

\eq\label{DYW_BD_quark}  
F_1^Q(Q^2) &=& \int\limits_0^1 dx \, 
\int \frac{d^2\bfk}{16 \pi^3} \, 
\biggl[ 
\psi^{\dagger\uparrow}_{Q; + \frac{1}{2}}(x,\bfk^{\prime+}) \,   
\psi^{\uparrow}_{Q; + \frac{1}{2}}(x,\bfk) 
+ 
\psi^{\dagger\uparrow}_{Q; - \frac{1}{2}}(x,\bfk^{\prime+}) \, 
\psi^{\uparrow}_{Q; - \frac{1}{2}}(x,\bfk) 
\biggr] \,, \\
F_2^Q(Q^2) &=& - \frac{2 M_N}{q^1-iq^2} \, 
\int\limits_0^1 dx \, 
\int \frac{d^2\bfk}{16 \pi^3} \, 
\biggl[ 
\psi^{\dagger\uparrow}_{Q; + \frac{1}{2}}(x,\bfk^{\prime+}) \,   
\psi^{\downarrow}_{Q; + \frac{1}{2}}(x,\bfk) 
+ 
\psi^{\dagger\uparrow}_{Q; - \frac{1}{2}}(x,\bfk^{\prime+}) \,  
\psi^{\downarrow}_{Q; - \frac{1}{2}}(x,\bfk) 
\biggr]
\en 
for heavy quark contributions and 
\eq\label{DYW_BD_antiquark}  
F_1^{\bar Q}(Q^2) &=& \int\limits_0^1 dx \, 
\int \frac{d^2\bfk}{16 \pi^3} \, 
\biggl[ 
\psi^{\dagger\uparrow}_{\bar Q; + \frac{1}{2}}(x,\bfk^{\prime-}) \,    
\psi^{\uparrow}_{\bar Q; + \frac{1}{2}}(x,\bfk) 
+ 
\psi^{\dagger\uparrow}_{\bar Q; - \frac{1}{2}}(x,\bfk^{\prime-}) \,     
\psi^{\uparrow}_{\bar Q; - \frac{1}{2}}(x,\bfk) 
\biggr] \,, \\
F_2^{\bar Q}(Q^2) &=& - \frac{2 M_N}{q^1-iq^2} \, 
\int\limits_0^1 dx \, 
\int \frac{d^2\bfk}{16 \pi^3} \, 
\biggl[ 
\psi^{\dagger\uparrow}_{\bar Q; + \frac{1}{2}}(x,\bfk^{\prime-}) \,     
\psi^{\downarrow}_{\bar Q; + \frac{1}{2}}(x,\bfk) 
+ 
\psi^{\dagger\uparrow}_{\bar Q; - \frac{1}{2}}(x,\bfk^{\prime-}) \,   
\psi^{\downarrow}_{\bar Q; - \frac{1}{2}}(x,\bfk) 
\biggr]
\en 
for the heavy antiquark contributions. Here, $\bfk^{\prime\pm} = \bfk \pm \bfq (1-x)$. 

A straightforward calculation gives 
the following expressions for the heavy quark-antiquark contributions 
to the proton Dirac and Pauli form factors:
\eq 
F_1^{Q-\bar Q}(Q^2) &=& F_1^Q(Q^2) - F_1^{\bar Q}(Q^2) \nonumber\\
&=& \int\limits_0^1 dx \, Q_{\rm asym}(x) \, 
\biggl[ 1 - \dfrac{Q^2 (1-x)^2}{8 \kappa^2}
\biggr] \, \exp\biggl[-\dfrac{Q^2 (1-x)^2}{4 \kappa^2}\biggr] \,, \\
\nonumber\\
F_2^{Q-\bar Q}(Q^2) &=& F_2^{Q}(Q^2) - F_2^{\bar Q}(Q^2) \nonumber\\
&=& \frac{M_N}{\kappa} \, \int\limits_0^1 dx \, 
(1-x) \, Q_{\rm asym}(x) 
\, \exp\biggl[-\dfrac{Q^2 (1-x)^2}{4 \kappa^2}\biggr]  \,.  
\en

One can see that the Dirac form factors are
properly normalized at $Q^2 = 0$: 
\eq 
F_1^Q(0) &=& \int\limits_0^1 dx \, Q_{\rm in}(x) = \frac{1}{100} 
\,, \qquad 
F_1^{\bar Q}(0) \ = \  
\int\limits_0^1 dx \, \bar Q_{\rm in}(x) = \frac{1}{100}
\,, \nonumber\\
F_1^{Q-\bar Q}(0) &=& F_1^Q(0) - F_1^{\bar Q}(0)
= \int\limits_0^1 dx \, Q_{\rm asym}(x) = 0 \,. 
\en 
The heavy quark-antiquark contributions to the nucleon Sachs form factors 
$G_{E/M}^{Q-\bar Q}(Q^2)$ and the electromagnetic radii
$\la r^2_{E/M}\ra^{Q-\bar Q}$ are given in terms of the Dirac and
Pauli form factors $F_{1,2}^{Q-\bar Q}(Q^2)$ as 
\eq 
G_{E}^{Q-\bar Q}(Q^2) &=& F_1^{Q-\bar Q}(Q^2) - \dfrac{Q^2}{4 M_N^2} 
\, F_2^{Q-\bar Q}(Q^2) \,, \nonumber\\
G_{M}^{Q-\bar Q}(Q^2) &=& F_1^{Q-\bar Q}(Q^2) + F_2^{Q-\bar Q}(Q^2) 
\,, \nonumber\\ 
\la r^2_{E}\ra^{Q-\bar Q} &=& 
- 6 \, \dfrac{G_{E}^{Q-\bar Q}(Q^2)}{dQ^2}\bigg|_{Q^2=0} 
\,, \nonumber\\
\la r^2_{M}\ra^{Q-\bar Q} &=& 
- \frac{6}{\mu_p} \, \dfrac{G_{M}^{Q-\bar Q}(Q^2)}{dQ^2}\bigg|_{Q^2=0} 
\,, 
\en  
where $G_M^{Q-\bar Q}(0) \equiv F_2^{Q-\bar Q}(0) = \mu^{Q-\bar Q}$ 
is the heavy quark-antiquark contribution to the proton magnetic moment. 
We have normalized the slope $\la r^2_{M}\ra^{Q-\bar Q}$ using  
the proton magnetic moment $\mu_p = 2.793$.
One can see that the magnetic moment $\mu^{Q-\bar Q}$ is proportional 
to the first moment of the asymmetry PDF $Q_{\rm asym}(x)$ and directly 
related to the first moment of the asymmetry distribution $\la x \ra_{Q-\bar Q}$ 
\eq 
\mu^{Q-\bar Q} &=& F_2^{Q-\bar Q}(0) = 
\frac{M_N}{\kappa} \, \int\limits_0^1 dx \, (1-x) \, Q_{\rm asym}(x)
\nonumber \\
&=& - \frac{M_N}{\kappa} \, \la x \ra_{Q-\bar Q} 
\en 

Our prediction for the charm-anticharm and bottom-antibottom contributions 
to the proton magnetic moment are negative, consistent with other 
theoretical predictions (see, e.g., discussion in Ref.~\cite{Sufian:2020coz}), 
and are equal to 
\eq 
\mu^{c-\bar c} = - 5.36 \times 10^{-4}\,, \qquad 
\mu^{b-\bar b} = - 4.94 \times 10^{-5} \,. 
\en 
It is clear why $\mu^{Q-\bar Q}$ is negative. The integrand is positive
and proportional to $(1-x)$. The $1$ contributes to the zero moment,
which is zero, while the $-x$ contributes to the first moment, leading to
the negative results for $\mu^{Q-\bar Q}$.  

We note that our prediction
for $\mu^{c-\bar c}$ is suppressed by a factor of 2.4 
in comparison with the result of Ref.~\cite{Sufian:2020coz}: 
$\mu^{c-\bar c} = - 1.27 \times 10^{-3}$. Our prediction for the relative
scaling of the charm-anticharm and bottom-antibottom contributions   
to the magnetic moment of proton is 
$\mu^{b-\bar b}/\mu^{c-\bar c} \sim (m_c/m_b)^2 \sim 0.1$.
Note that a negative contribution to the magnetic
moment of the heavy quarks is also a property of the strange quark
contribution.
A detailed discussion is given in Ref.~\cite{Lyubovitskij:2002ng}
in the framework of the perturbative chiral quark model (PCQM).
The predictions of the PCQM for the strange quark contribution
to the nucleon properties have been successfully confirmed from
data~\cite{Maas:2017snj} and lattice
calculations~\cite{Green:2015wqa,Alexandrou:2019olr}.

Analytic result for the intrinsic heavy quark contributions
to the electromagnetic radii of the proton can be obtained using 
derivatives of the Dirac and Pauli form factors: 
\eq 
F_1^{\prime; Q-\bar Q}(0) &=& - \frac{3}{8 \kappa^2} \, \int\limits_0^1 dx \, 
(1-x)^2 \,  Q_{\rm asym}(x) \,, \nonumber\\
F_2^{\prime; Q-\bar Q}(0) &=& - \frac{M_N}{4 \kappa^3} \, \int\limits_0^1 dx \, 
(1-x)^3 \,  Q_{\rm asym}(x) \,. 
\en 

\begin{figure}
\begin{center}
\epsfig{figure=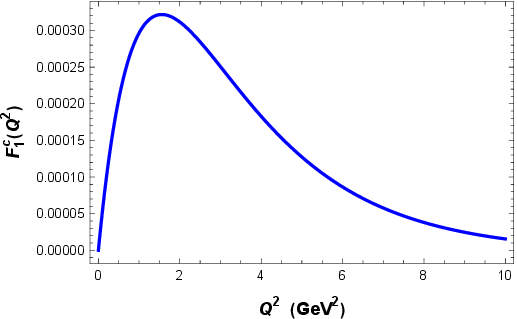,scale=.75}
\hspace*{1cm}
\epsfig{figure=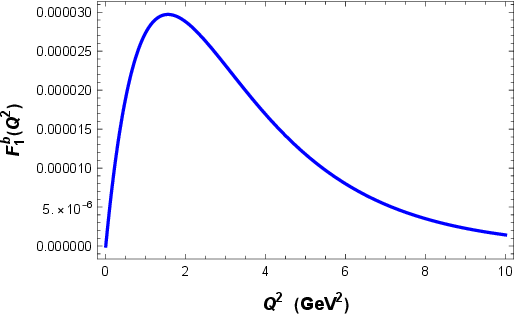,scale=.75}
\end{center}
\vspace*{-.48cm}
\noindent
\caption{Heavy quark-antiquark contributions to the 
  Dirac form factor of proton.
  Left panel: charm-anticharm. Right panel: bottom-antibottom. 
\label{fig4}}

\begin{center}
\epsfig{figure=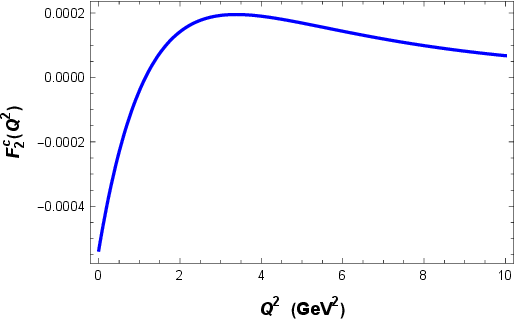,scale=.75}
\hspace*{1cm}
\epsfig{figure=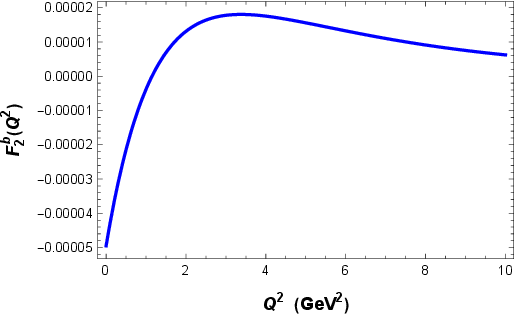,scale=.75}
\end{center}
\vspace*{-.48cm}
\noindent
\caption{Heavy quark-antiquark contributions to the 
  Pauli form factor of proton. Left panel: charm-anticharm.
  Right panel: bottom-antibottom. 
\label{fig5}}

\begin{center}
\epsfig{figure=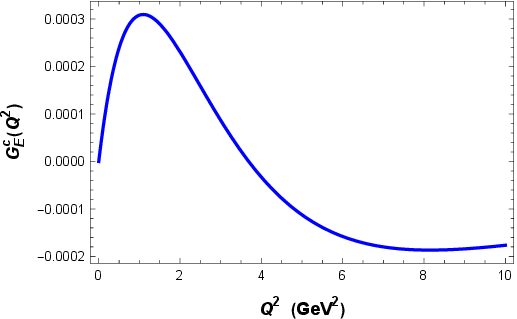,scale=.75}
\hspace*{1cm}
\epsfig{figure=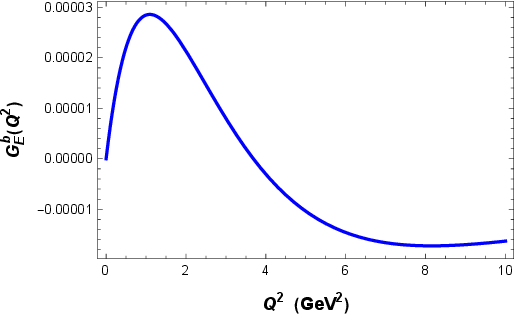,scale=.75}
\end{center}
\vspace*{-.48cm}
\noindent
\caption{Heavy quark-antiquark contributions to the 
  Sachs charge form factor of proton. Left panel: charm-anticharm.
  Right panel: bottom-antibottom. 
\label{fig6}}

\begin{center}
\epsfig{figure=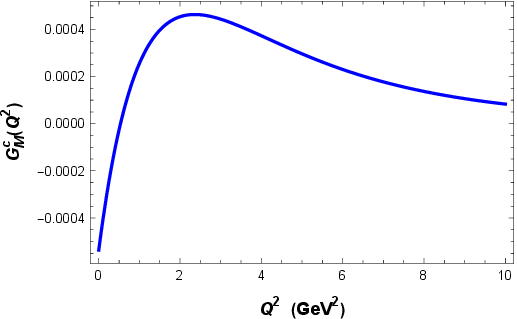,scale=.75}
\hspace*{1cm}
\epsfig{figure=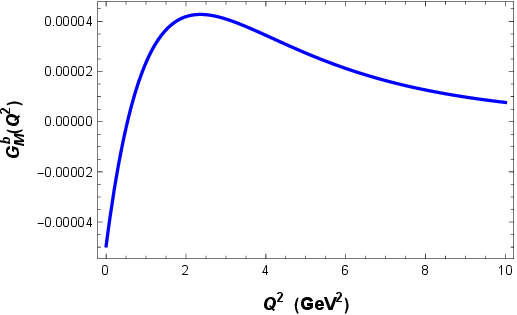,scale=.75}
\end{center}
\vspace*{-.48cm}
\noindent
\caption{Heavy quark-antiquark contributions to the 
  Sachs magnetic form factor of proton. Left panel: charm-anticharm.
  Right panel: bottom-antibottom. 
\label{fig7}}
\end{figure}

\begin{figure}
\begin{center}
\epsfig{figure=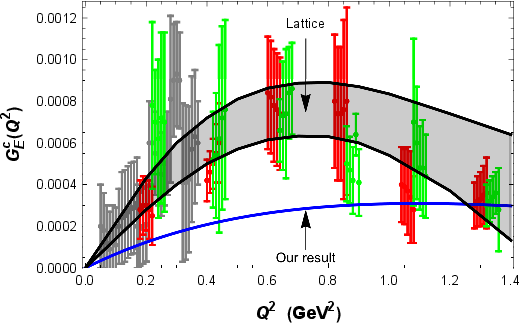,scale=.75}
\hspace*{1cm}
\epsfig{figure=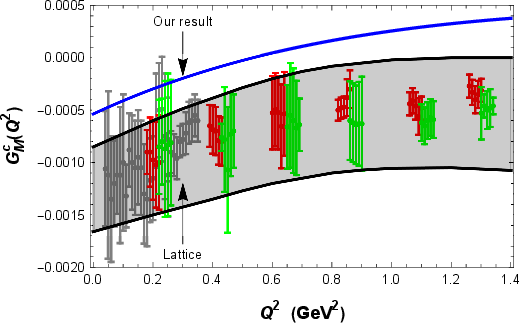,scale=.75}
\end{center}
\vspace*{-.48cm}
\noindent
\caption{Comparison of our results for the charm-anticharm
  contributions to the Sachs form factors
  $G_E^c(Q^2)$ (left panel) and $G_M^c(Q^2)$ (right panel)
  with lattice predictions~\cite{Sufian:2020coz}
  (shaded band and data points with errors). 
\label{fig8}}
\end{figure}

One gets: 
\eq 
\la r^2_{E}\ra^{Q-\bar Q} &=& 
- 6 \, F_1^{\prime; Q-\bar Q}(0) + \dfrac{3}{2 M_N^2} \, 
\mu^{Q-\bar Q} 
\,, \nonumber\\
\la r^2_{M}\ra^{Q-\bar Q} &=& 
- 6 \, \biggl[F_1^{\prime; Q-\bar Q}(0) + F_2^{\prime; Q-\bar Q}(0) 
\biggr] \,. 
\en 
Numerical results for the charm-anticharm and bottom-antibottom 
contributions to the electromagnetic slopes are: 
\eq
& &\la r^2_{E}\ra^{c-\bar c} = -1.70 \times 10^{-4} \, {\rm fm}^2 \,,
\qquad 
\la r^2_{M}\ra^{c-\bar c} = -1.12 \times 10^{-4} \, {\rm fm}^2 \,,
\nonumber\\
& &\la r^2_{E}\ra^{b-\bar b} = -1.57 \times 10^{-5} \, {\rm fm}^2 \,,
\qquad 
\la r^2_{M}\ra^{b-\bar b} = -1.03 \times 10^{-5} \, {\rm fm}^2 \,.
\en 

Notice that in our approach the following scaling laws for the 
ratios of the charm-anticharm and bottom-antibottom contribution
to the proton radii are: 
$\la r^2_{E,M}\ra^{b-\bar b}/\la r^2_{E,M}\ra^{c-\bar c} \sim (m_c/m_b)^2 \sim 0.1$. 
We also note that our results for the charm sector are in
agreement with the results of Ref.~\cite{Sufian:2020coz} 
by the same sign (negative) and in order of magnitude:  
$\la r^2_{E}\ra^{c-\bar c} = - 0.0005(1)$ fm$^2$ and 
$\la r^2_{M}\ra^{c-\bar c} = - 0.0003(1)$ fm$^2$.  
We plot our results for the the
charm-anticharm and bottom-antibottom contributions 
to the Dirac, Pauli, and Sachs form factors in Figs.~\ref{fig4}-\ref{fig8}.  
In particular, in Fig.~\ref{fig8} we present a comparison of our results
for the Sachs form factors
$G_E^c(Q^2)$ (left panel) and $G_M^c(Q^2)$ (right panel)
form factors with lattice predictions~\cite{Sufian:2020coz}
(shaded band and data points with errors). 

An important check of our approach concerns the model-independent 
result obtained in Ref.~\cite{Brodsky:2000ii} for the vanishing
of the anomalous gravitomagnetic moment $B(0) = 0$ for each Fock state
contributing to a composite system, consistent with the general theorem
of Okun and Kobzarev~\cite{Kobzarev:1962wt}. 
In particular, we will verify 
that the contribution of the Fock state describing a presence of
the heavy quark-antiquark in the proton to the $B(0)$ is zero.

Using the master formula for the $B(0)$ derived in Ref.~\cite{Brodsky:2000ii}, 
one has 
\eq
- \dfrac{B_{Q}(0)}{2 M_N} &=&  \lim\limits_{q_\perp^1 \to 0} \,
\dfrac{\partial}{\partial q_\perp^1}  \,
\int\limits_0^1 dx \, \int \frac{d^2\bfk}{16 \pi^3} \  
\sum\limits_{\lambda_Q = \pm \frac{1}{2}} \biggl[
x \, \psi^{\dagger\uparrow}_{Q; \lambda_Q}(x,\bfk^{\prime+}) \, 
\psi^{\downarrow}_{Q; \lambda_Q}(x,\bfk) +
(1-x) \, \psi^{\dagger\uparrow}_{Q; {\lambda_Q}}(x,\bfk^{\prime\prime+}) \, 
\psi^{\downarrow}_{Q; \lambda_Q}(x,\bfk) \biggr] 
\nonumber\\
&=&
\lim\limits_{q_\perp^1 \to 0} \, 
\dfrac{\partial}{\partial q_\perp^1} \, 
\biggl[ \dfrac{q_\perp^1}{\kappa} \, 
\int\limits_0^1 dx \ Q_{\rm in}(x) \ x (1-x) 
\ \underbrace{(1 - 1)}_{= 0} \ 
\exp\Big[-\dfrac{(q_\perp^1)^2}{\kappa^2}\Big]  \biggr] = 0 
\en
for the struck heavy quark and
\eq
- \dfrac{B_{\bar Q}(0)}{2 M_N} &=&
\lim\limits_{q_\perp^1 \to 0} \,
\dfrac{\partial}{\partial q_\perp^1}  \,
\int\limits_0^1 dx \, \int \frac{d^2\bfk}{16 \pi^3} \ 
\sum\limits_{\lambda_{\bar Q} = \pm \frac{1}{2}} \biggl[
x \, \psi^{\dagger\uparrow}_{Q; \lambda_{\bar Q}}(x,\bfk^{\prime-}) \, 
\psi^{\downarrow}_{Q; \lambda_{\bar Q}}(x,\bfk) +
(1-x) \, \psi^{\dagger\uparrow}_{Q; \lambda_{\bar Q}}(x,\bfk^{\prime\prime-}) \, 
\psi^{\downarrow}_{Q; \lambda_{\bar Q}}(x,\bfk) \biggr] 
\nonumber\\
&=&
\lim\limits_{q_\perp^1 \to 0} \,
\dfrac{\partial}{\partial q_\perp^1}  \, 
\biggl[ \dfrac{q_\perp^1}{\kappa} \, 
\int\limits_0^1 dx \ \dfrac{7}{5} \ Q_{\rm in}(x) \
x (1-x)^2 \ \underbrace{(1 - 1)}_{= 0} \
\exp\Big[-\dfrac{(q_\perp^1)^2}{\kappa^2}\Big] \biggr] = 0 
\en 
for the struck heavy antiquark. Here
$\bfk^{\prime\pm} = (k^1_\perp \pm q^1_\perp (1-x), k^2_\perp)$ and 
$\bfk^{\prime\prime\pm}= (k^1_\perp \mp  q^1_\perp x, k^2_\perp)$. 

In conclusion, we derived LFWFs describing heavy quark-antiquark
content in the proton. 
Using these LFWFs, we construct light-front representations for the
proton properties induced by heavy quarks and
made numerical applications.

\begin{acknowledgments}

This work was funded by BMBF (Germany) ``Verbundprojekt 05P2021 (ErUM-FSP T01) -
Run 3 von ALICE am LHC: Perturbative Berechnungen von Wirkungsquerschnitten
f\"ur ALICE'' (F\"orderkennzeichen: 05P21VTCAA), by ANID PIA/APOYO AFB180002 (Chile),
by FONDECYT (Chile) under Grant No. 1191103
and by ANID$-$Millen\-nium Program$-$ICN2019\_044 (Chile).
The work of SJB was supported in part by the Department of Energy
under contract DE-AC02-76SF00515. SLAC-PUB-17698.

\end{acknowledgments}

\end{document}